\documentclass[prb,preprint,groupedaddress,showpacs,showemail]{revtex4}
\usepackage{graphicx}

\begin{document}

\title{Properties of low-lying excited manifolds 
in the Mn$_{12}$ acetate}
\author{Kyungwha Park$^{1,2}$}\email{park@dave.nrl.navy.mil}
\author{Mark R. Pederson$^{1}$}\email{pederson@dave.nrl.navy.mil}
\author{C.~Stephen Hellberg$^{1}$}\email{hellberg@dave.nrl.navy.mil}
\affiliation{
$^1$Center for Computational Materials Science, Code 6390,
Naval Research Laboratory, Washington DC 20375 \\
$^2$Department of Electrical Engineering and Materials Science Research Center,
Howard University, Washington DC 20059}
\date{\today}

\begin{abstract}
Most experimental data on the single-molecule magnet Mn$_{12}$ acetate
have been successfully explained by the assumption that 
the Mn$_{12}$ acetate has an effective ground-state spin of $S=10$. However, 
the effect of the low-lying excited manifolds caused by interactions
between Mn spins has not been well understood. To investigate the features of
the low-lying excited manifolds, the intramolecular exchange interactions are 
calculated using density-functional theory (DFT). With the calculated
exchange parameters, the energy gap between the $S=10$ ground-state 
and the first excited-state manifold is calculated by diagonalization of
the Heisenberg Hamiltonian. The upper limit on the energy gap is about 40.5~K 
which is likely to be overestimated due to incomplete treatment of the Coulomb 
potential within DFT. It is found that there are several $S=9$ low-energy 
excited-state manifolds above the $S=10$ ground-state manifold. 
The magnetic anisotropy barriers for the low-lying spin excitations are calculated
using DFT. Based on the calculations, it is found that the anisotropy barriers for the
low-lying excited manifolds are approximately the same as that for the
ground-state manifold, which is applicable to other single-molecule magnets such as Mn$_4$.
\end{abstract}

\maketitle


\section{Introduction}

\begin{figure}
\includegraphics[angle=0,width=0.4\textwidth]{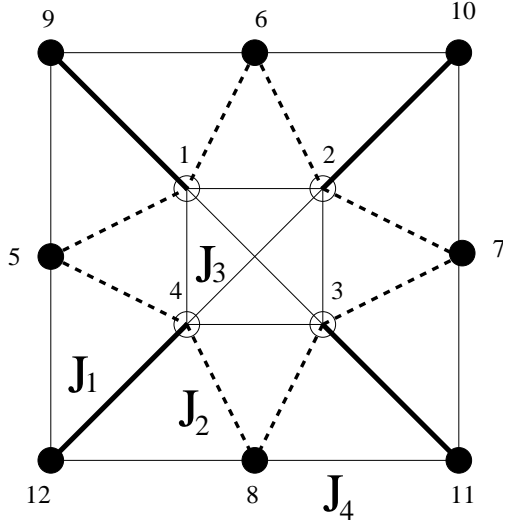}
\caption{Schematic diagram of the exchange interactions within
a single molecule for the SMM Mn$_{12}$. The empty circles
denote Mn$^{4+}$ ions, the filled circles denote Mn$^{3+}$ ions,
and each Mn ion is numerically labeled. The bond length between
spin 1 and 9 is shorter than that between
spin 1 and 6. The thick solid lines are for $J_1$, the thick 
dashed lines are for $J_2$, the thin solid lines in the inner cubane
are for $J_3$, and the thin solid lines in the outer crown
are for $J_4$.}
\label{fig:mn12sch}
\end{figure}

Single-molecule magnets (SMMs) have been extensively studied for 
the past decade because of both scientific and practical reasons: 
macroscopic quantum phenomena\cite{CHUD98} and possible utilization 
as magnetic storage devices.\cite{JOAC00} A prototype of the SMMs is 
[Mn$_{12}$O$_{12}$(CH$_3$COO)$_{16}$(H$_2$O)$_4$]
$\cdot$2(CH$_3$COOH)$\cdot$4(H$_2$O) (hereafter
Mn$_{12}$),\cite{LIS80} which is a three-dimensional array of identical
$S=10$ molecules. Very recently, derivatized Mn$_{12}$ type molecules 
were successfully deposited on a gold film,\cite{CORN03} which 
further enhances the prospects for storing magnetic information 
in a single molecule. 
A single molecule of Mn$_{12}$ has four ferromagnetically coupled
Mn$^{4+}$ ions ($S=3/2$) in the cubane and eight ferromagnetically
coupled Mn$^{3+}$ ions ($S=2$) in the crown as schematically 
shown in Fig.~\ref{fig:mn12sch}. In the ground state, the magnetic 
moments of the eight Mn$^{3+}$ ions are antiparallel to those 
of the four Mn$^{4+}$ ions which leads to a total spin of $S=10$. 

So far many interesting features of the SMM Mn$_{12}$ such as
magnetization steps in the hysteresis loops,\cite{SESS93-NAT,FRIE96} electron 
paramagnetic resonance (EPR) transitions,\cite{BARR97,HILL98} and 
low-energy excitations in inelastic neutron scattering,\cite{HENN97,CACI98,MIRE99} 
have been well understood by considering each molecule as an 
object with an effective spin of $S=10$. 
However, magnetic susceptibility measurements\cite{CANE92,MUKH98,GOME98} 
have demonstrated that some experimental data could not be in accord with the $S=10$ 
single-spin picture and suggested that the first excited-state manifold 
may be located at 35~K$-$40~K above the $S=10$ ground-state manifold 
within which the energy barrier to magnetization reversal is 65~K in zero field.
Since the first excited-state manifold overlaps with the ground-state
manifold above 35~K$-$40~K, an internal many-spin structure
of a single molecule should be included to explain the high-energy
experimental data. The dimension of the Hilbert space of this SMM is 
so large that a simplified eight-spin model with strong Dzyaloshinsky-Morya 
interaction was first proposed to include many-spin effects 
by dimerizing two strongly bonded Mn spins.\cite{KATS99} 
Although this eight-spin model explained some experimental data,
it was limited to features below 50~K. At higher temperatures 
the dimerization scheme breaks down. Later all twelve Mn spins were
included in the Heisenberg Hamiltonian and excited-state manifolds
were clarified by diagonalization of the Hamiltonian with constraining 
the fixed energy gap between the first excited-state and the ground-state 
manifold of 35~K.\cite{RAGH01,REGN02} Nonetheless,
there is still a big controversy over the energy gap between the 
first excited-state and the ground-state manifold, and 
there has been less exploration of high-energy features caused by 
many-spin effects. Here ``high-energy''
means low-lying excited manifolds above the $S=10$ ground-state 
manifold. Recent high-energy inelastic neutron scattering 
measurements exhibited a broad anomalous peak at 10~K, which may have 
a magnetic origin (not due to phonons) and could not be rationalized 
by the $S=10$ single-spin picture because the first allowed peak
by the picture is about 14~K.\cite{HENN97} More recent high-field EPR 
measurements revealed that several transitions could not be
justified by the $S=10$ single-spin manifold but that they were well
explained if we assumed the $S=9$ first excited-state manifold to be 
situated at 10-16~K above the ground-state manifold.\cite{EDWA03} 
Additionally, there have been many attempts to determine the signs 
and magnitudes of the exchange couplings between Mn ions from different
approaches\cite{SESS93,RAGH01,BOUK02,REGN02} but no consensus
has arisen. 

To elucidate the energy gap and examine properties
within the low-energy excited manifolds, a single molecule
of the Mn$_{12}$ is considered and the exchange interactions 
between Mn spins are calculated using density-functional theory (DFT), 
which are presented in Secs.~II and III. Our calculated values
are compared with those from other groups.\cite{SESS93,RAGH01,BOUK02,REGN02}
Through diagonalization of the isotropic Heisenberg exchange Hamiltonian
with our calculated exchange constants using the 
Lanczos method,\cite{CULL85,HELL99}
excited-state manifolds are manifested and energy gaps between
different manifolds are calculated in Sec.~IV. The magnetic anisotropy 
barriers for low-energy spin excitations are calculated
using DFT and the anisotropy barriers for low-lying excited manifolds
are discussed in Sec.~V. Our conclusions are presented in Sec.~VI.

\section{DFT calculations}

Our DFT calculations\cite{KOHN65} are performed with
spin-polarized all-electron Gaussian-orbital-based Naval Research
Laboratory Molecular Orbital Library (NRLMOL) \cite{PEDE90} which
is ideal for studying a single molecule to a small number of unit
cells. Here we use the Perdew-Burke-Ernzerhof (PBE)
generalized-gradient approximation (GGA) in the
exchange-correlation potential.\cite{PERD96} Since the SMM Mn$_{12}$ 
has fourfold symmetry, for fully symmetrized calculations the total 
number of inequivalent atoms to consider is reduced to $176/4=44$.
To save geometry-optimization time without losing interesting
physical properties, the following simplified form of the SMM
Mn$_{12}$ is used: [Mn$_{12}$O$_{12}$(HCOO)$_{16}$(H$_2$O)$_4$]
(acetic acids and water molecules of crystallization are not included, 
and 16 acetates, CH$_3$COO, are replaced by 16 formates, HCOO). 
The zero-field total anisotropy barrier for the $S=10$ ground-state
manifold does not change much with this simplification. The total
magnetic moment for the ground state was confirmed to be 
$20 \mu_{\mathrm B}$, which is in good agreement with experiment 
and it is stable. Details of the optimization schemes and electronic
properties of the optimized geometry for the ground state were
discussed elsewhere.\cite{PEDE99} Hereafter, unless specified,
our calculations have been carried out for the above simplified form.

\section{Intramolecular exchange interactions}

The fourfold symmetry and the geometry of a single Mn$_{12}$ molecule 
indicate that there are three symmetrically inequivalent 
Mn sites and four different exchange interactions between Mn spins 
as shown in Fig.~\ref{fig:mn12sch}. To calculate the exchange
constants using DFT, it is assumed that the magnetic moments of
all Mn ions are aligned along a particular direction (collinear). 
Then eleven distinctive spin configurations 
are constructed by reversing magnetic moments of a few Mn ions
simultaneously from the $S=10$ ground-state. (See Table~\ref{table:1}) 
For example, a $M_s=9$ state ($M_s$ is an eigenvalue of the 
total spin operator $S$ projected along the particular direction) 
can be built by flipping both one Mn spin in the cubane and another 
Mn spin in the crown within the $S=10$ ground state: 
$M_s=-3/2 \times (3-1)+2 \times (7-1)=9$. There are several ways 
to construct $M_s=9$ states with collinear spins that are
labeled as 9-b, 9-c, and 9-d in Table~\ref{table:1}. 
All of the examined collinear spin configurations are {\it not} eigenstates 
of $S^2$. Mean-field calculations are not directly applicable to 
optimizing excited states such as $|S=9,M_s=9 \rangle$ in Mn$_{12}$, 
because those eigenstates are
represented by linear combinations of many single Slater determinants.
The geometries for the distinctive spin configurations are taken
to be the same as that for the ground state except for the proper
spin arrangements. It is confirmed that a slightly different initial geometry
for the collinear spin configurations does not siginificantly alter 
our calculated values of the exchange constants. 
Since fourfold symmetry is broken for the  
spin configurations considered, unsymmetrized calculations are required
with a total number of 100 atoms. The energies of the spin configurations
are self-consistently minimized with a small basis set and fine mesh.\cite{PORE99} 
Our calculations show that the minimized energies range from 0.04~eV to 0.2~eV
above the ground-state ferrimagnetic structure.
From Ref.~\onlinecite{PEDE99} it is known that for the $S=10$ ground state 
the lowest electronic excitation is a majority spin excitation 
between the $e_g$ levels in Mn(3d) states, which is about 0.44~eV
(majority LUMO-HOMO gap),
and that the energy gap between minority lowest unoccupied molecular
orbital (LUMO) and majority highest occupied molecular orbital (HOMO) 
is $\sim 0.89$~eV. Comparison of the minimized energies
to the LUMO-HOMO gaps indicates that these different spin 
configurations are lower-energy spin excitations than moving
an electron from the majority HOMO to the unoccupied majority (or minority) 
orbital. Thus they will be called low-energy spin excitations.
There are twelve equations to solve for five unknowns
(the background energy, $E_0$, four exchange constants 
$J_1$, $J_2$, $J_3$, and $J_4$) so a least-square-fit (LSF) 
method\cite{NUME} is used. 

Our calculated values of $J$'s are shown in Table~\ref{table:2} in
comparison with results of other groups. When the minimized energies of
the low-energy spin excitations are recalculated using the
calculated values of $J$'s, it is found that they are in 
good agreement with the DFT-calculated energies as indicated
in Table~\ref{table:1}. Prior to comparison to the results of other groups,
let us briefly review the essence of the different approaches from 
the different groups. 
In Ref.~\onlinecite{SESS93} the value of $J_1$ was first determined
from the experimental data, and then $J_2$ and $J_3$ were varied
with fixed value of $J_4=0$ to reproduce the $S=10$ ground state.
In Ref.~\onlinecite{RAGH01} the values of $J$'s were slightly 
varied from the previously reported values\cite{SESS93}
using diagonalization of Heisenberg exchange Hamiltonian
to provide the $S=10$ ground state and the energy gap of 35~K
between the ground-state and the $S=9$ first excited-state manifold.
In Ref.~\onlinecite{REGN02} the exchange parameters were obtained
to reproduce their megagauss experimental data with a constraint of
the energy gap of 35~K. In Ref.~\onlinecite{BOUK02} the LDA$+$U method was used
to include electron correlations between Mn ions, and the values of
the intra-atomic Hund's exchange parameter $J$ and the average Coulomb 
parameter $U$ were determined to obtain the correct magnetic moment 
for the $S=10$ ground state. Then the exchange constants were calculated from 
the values of $J$ and $U$. Our calculations show that
the coupling between one Mn spin in the 
cubane and another Mn spin in the crown closest to the Mn spin in the
cubane, $J_1$, is confirmed to be the strongest antiferromagnetic 
coupling, although its absolute magnitude does not agree among 
different groups. Our calculated value of $J_1$ agrees somewhat 
with those from Refs.~\onlinecite{BOUK02,REGN02}. The coupling
between a Mn spin in the cubane and the other type of Mn spin
in the crown, $J_2$, is confirmed to be weaker antiferromagnetic 
than $J_1$. But the rate of $J_2$ to $J_1$ 
does not agree among different groups. Our value of $J_2$ is
between the values obtained from Refs.~\onlinecite{BOUK02,REGN02}. 
The coupling between Mn spins in the cubane, $J_3$, turns out to be 
the weakest and is ferromagnetic. 
The coupling between Mn spins in the crown, $J_4$, is weakly
antiferromagnetic. Overall, the DFT calculated values are
closest to the results obtained from Ref.~\onlinecite{REGN02},
and they are 20\%-40\% larger than those from the LDA$+$U method
except for the weakest $J_3$. 
However, they are significantly smaller than the other
two reported results from Refs.~\onlinecite{RAGH01,SESS93}.
Our calculated values of $J$' may be overestimated compared to
experimental data for the following reasons.
It has been found from other SMMs such as Mn$_4$[Ref.~\onlinecite{HEND92}] 
and V$_{15}$[Ref.~\onlinecite{MULL88}] that the DFT-calculated exchange 
interactions were overestimated by a factor of 2 and 3 compared to the 
experimental data respectively.\cite{PARK03-2,KORT01} 
This overestimated exchange interactions are
due to the fact that the PBE generalized-gradient
approximation does not fully treat self-interaction
corrections of the localized $d$ states in the Coulomb potential.
Proper treatment of the electron correlations in Mn atoms
may improve our DFT calculations of the exchange interactions.
This is supported by the fact that the LDA$+$U method produces 
20\%-40\% smaller exchange interactions than our DFT calculations 
and that the exchange interactions decrease with increasing the value
of $U$ in the LDA$+$U method.\cite{BOUK02}

\section{Heisenberg exchange Hamiltonian}

To calculate the energy gap between the ground-state
and the first excited manifold,
we use the following isotropic (i.e. no anisotropy) 
Heisenberg exchange Hamiltonian:
\begin{equation}
{\cal H} = \sum_{i < j} J_{ij} \vec{S}_i \cdot \vec{S}_j
\label{eq:ham_ms}
\end{equation}
where the sum runs over all pairs connected by the exchange interactions
and the values of the coupling constants $J_{ij}$ labeled 
in Fig.\ref{fig:mn12sch} are in Table~\ref{table:2}. 
The exchange Hamiltonian is a $10^8 \times 10^8$ matrix, which is 
too large to diagonalize using currently available computers. 
Thus, using the fact that $M_s$ is a good quantum number, 
we classify the total number of the $10^8$ states into 45 different 
constant $M_s$ states such as 
$M_s=\pm 22, \pm 21, \pm 20,..., \pm 10, \pm 9,..., \pm 1, 0$.
Since we are interested in low-lying excited manifolds, 
we examine $M_s=-12$, $M_s=-11$, $M_s=-10$, $M_s=-9$, and $M_s=-8$ states only. 
If the $S=10$ manifold is the ground state, then the lowest-energy 
eigenstate of the exchange Hamiltonian with a basis set of 
all $M_s=-10$ states corresponds to $|S=10, M_s=-10 \rangle$. 
Additionally the lowest-energy eigenstate of the Hamiltonian with a basis 
set of all $M_s=-9$ states corresponds to $|S=10,M_s=-9 \rangle$
and it should have the same energy as $|S=10, M_s=-10 \rangle$. 
The eigenstates with the same $S$ must be degenerate 
if they belong to the same manifold. The first-excited eigenstate of the
Hamiltonian with the basis set of all $M_s=-9$ states corresponds 
to $|S=9,M_s=-9 \rangle$ if the $S=9$ manifold is the first-excited
manifold above the $S=10$ ground state.
There are a total number of 269148 $M_s=12$ states, 
484144 $M_s=-11$ states, 817176 $M_s=-10$ states, and 1299632
$M_s=-9$ states, 1954108 $M_s=-8$ states.

\begin{figure}
\includegraphics[angle=0,width=0.4\textwidth]{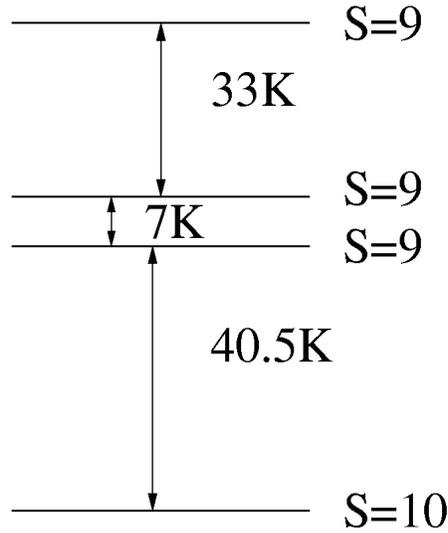}
\caption{Calculated low-lying excited manifolds for the SMM Mn$_{12}$.}
\label{fig:Elevel}
\end{figure}

We diagonalize the full Hamiltonian matrices for the $M_s=-12$, 
$M_s=-11$, $M_s=-10$, $M_s=-9$, and $M_s=-8$ states using the Lanczos 
method.\cite{CULL85,HELL99} A few calculated low-lying energy eigenvalues 
are shown in Table~\ref{table:3}. The lowest energy for the $M_s=-10$ 
states is identical to those for the $M_s=-9$ and $M_s=-8$ states, but 
it is much lower than the lowest energies for the $M_s=-11$ and $M_s=-12$ states.
The lowest energies for the $M_s < -12$ states will increase with increasing
$|M_s|$. This indicates that the ground-state has $S=10$. Next higher energies are 
from the $M_s=-9$ and $M_s=-8$ states. The three excited energy eigenvalues 
for the $M_s=-9$ states coincide with those for the $M_s=-8$ states 
which are much lower than the excited energies for the $M_s=-10$, $M_s=-11$, 
and $M_s=-12$ states. This suggests that the three excited states
for the $M_s=-9$ ($M_s=-8$) states belong to $S=9$ manifolds. 
So the first excited-state manifold has $S=9$ 
located at 40.5~K above the $S=10$ ground state, 
and there are two more $S=9$ excited manifolds located at 7~K and 40~K
above the $S=9$ first excited manifold (Fig.~\ref{fig:Elevel}). 
Our calculated values of the
energy gaps are upper limits because DFT calculations may overestimate
the exchange interactions due to the reasons discussed in Sec.~III.
If the exchange constants are reduced by a half, then the energy gaps
will be also reduced by a half. Single-ion anisotropy parameters which
we did not include in Eq.~(\ref{eq:ham_ms}) does not significantly affect 
the energy gap.

\section{Magnetic anisotropy barrier}

The magnetic anisotropy barriers for the eleven
low-energy spin excitations as well as the ground state
are calculated with the assumption that the barriers are 
caused by the spin-orbit interaction only. Other effects 
such as noncollinearity of the magnetic moments of Mn ions 
and spin-orbit-vibron coupling\cite{PEDE02} on the anisotropy barrier 
will not be included in our study. However correlation effects due to the
addition of multideterminantal wavefunctions are addressed in part by the 
further ongoing discussion.
The spin-orbit interaction $V_{LS}$ can be decomposed into one-electron 
operators $f_i$ related to the electric fields caused by nuclei and 
two-electron operators $g_{ij}$ related to the electric fields
due to the rest of electrons as follows:
\begin{eqnarray}
V_{LS}&=& -\frac{1}{2c^2} \sum_i \vec{S}_i \cdot
(\vec{p}_i \times \vec{\nabla} \Phi_{i} )
\label{eq:vls-1} \\
      &=& \sum_i f_i + \sum_{i \neq j} g_{ij}~,
\label{eq:vls-2} \\
\Phi_i &=& 
\sum_{\nu} \frac{Z_{\nu}}{|\vec{r}_i - \vec{R}_{\nu}|}
+ \sum_{j\neq i} \frac{1}{|\vec{r}_i - \vec{r}_j|} \label{eq:vls-3} \\
 f_i&=& \frac{1}{i} \sum_{\nu} \frac{Z_{\nu}}{2c^2} \vec{S}_i \cdot 
\left( \vec{\nabla}_i \times  \frac{(\vec{r}_i - \vec{R}_{\nu})}
{|\vec{r}_i - \vec{R}_{\nu}|^3} \right) \label{eq:vls-4} \\
g_{ij}&=&\frac{1}{i}\frac{1}{2c^2} \vec{S}_i \cdot 
\left( \vec{\nabla}_i \times 
\frac{(\vec{r}_i - \vec{r}_j)}{|\vec{r}_i - \vec{r}_j|^3} \right)
\label{eq:vls-5}
\end{eqnarray}
where the summation for $i$ ($\nu$) runs over all electrons 
located at $\vec{r}_i$ (all nuclei located at $R_{\nu}$ with nuclear number $Z_{\nu}$), 
$c$ is the speed of light, $\vec{S}_i$ is the spin operator 
of the $i$th electron, and $\vec{p}_i$ is the momentum of 
the $i$th electron. Within the self-consistent field (SCF) approximation, 
we have calculated the wave function 
$\Psi_{i \sigma}=\psi_{i \sigma}(\vec{r}) \chi_{\sigma}$ satisfying 
${\cal H}|\Psi_{i \sigma} \rangle = \epsilon_{i\sigma} | \Psi_{i \sigma}\rangle$, 
where ${\cal H}$ is the many-electron Hamiltonian
without spin-orbit coupling, $\psi_{i \sigma}$ is a spatial
function and $\chi_{\sigma}$ is a spinor. Therefore, the energy shift,
$\Delta_2$, due to $V_{LS}$ using second-order perturbation theory
is written by\cite{PEDE99}
\begin{eqnarray}
\Delta_2 &=& \sum_{\sigma, \sigma^{\prime}} \sum_{ij}
\frac{ \langle \Psi_{i \sigma} | V_{LS} | \Psi_{j \sigma^{\prime}} \rangle
\langle \Psi_{j \sigma^{\prime}} | V_{LS} | \Psi_{i \sigma} \rangle }
{\epsilon_{i \sigma} - \epsilon_{j \sigma^{\prime}}}
\label{eq:delta2}
\end{eqnarray}
where $\epsilon_{i \sigma}$ is the energy of the occupied state
with spinor ${\sigma}$, and $\epsilon_{j \sigma^{\prime}}$ is the
energy of the unoccupied state with spinor ${\sigma^{\prime}}$.
The summation runs over all occupied and unoccupied states of all
atoms within a certain energy window and over up and down spinor
states. The contributions for the cases that both of $i$ and $j$
are occupied (or unoccupied) to the energy shift cancel
each other out. For uniaxial systems such as Mn$_{12}$ 
the second-order energy shift is simplified as follows:
\begin{eqnarray}
\Delta_2 &=& D \langle S_z \rangle^2
\label{eq:gamma}
\end{eqnarray}
where $D$ is the uniaxial anisotropy parameter which can be calculated
from Eq.~(\ref{eq:delta2}). For the $S=10$ ground state, 
as shown in Table~\ref{table:1} the uniaxial parameter is about 0.54~K,
which is in good agreement with experiment.\cite{BARR97}

An eigenstate $|\Phi \rangle$ of the many-electron Hamiltonian 
that belongs to an excited-state manifold can be written 
in terms of a linear combination of many Slater determinants: 
$|\Phi \rangle = \sum_{\mu} C_{\mu} | \Phi_{\mu} \rangle$, 
where $|\Phi_{\mu} \rangle$ is a single Slater determinant 
(orthonormal basis function), and some of $|\Phi_{\mu} \rangle$ 
correspond to our examined low-energy spin excitations.
Therefore, the expectation value of the spin-orbit coupling 
$V_{LS}$ with respect to $|\Phi \rangle$, is decomposed into
diagonal and off-diagonal elements such as
\begin{eqnarray}
\langle \Phi | V_{LS} | \Phi \rangle &=&
\sum_{\mu} C^{\ast}_{\mu} C_{\mu} \langle \Phi_{\mu} |V_{LS}|
\Phi_{\mu}  \rangle + \sum_{\mu \neq \nu} C^{\ast}_{\mu} C_{\nu}
\langle \Phi_{\mu} |V_{LS}| \Phi_{\nu}  \rangle \;,
\label{eq:vls-6} 
\end{eqnarray}
where the summations run over spatial 
coordinates and spin variables of all electrons in Mn$_{12}$. 
$V_{LS}$ is made of one-electron and two-electron operators as shown in
Eq.~(\ref{eq:vls-2}). To facilitate calculations of the diagonal 
and off-diagonal elements, the off-diagonal elements of the one-electron
operator, $\sum_i \langle \Phi_{\mu} |f_i| \Phi_{\nu} \rangle$, 
are explicitly written in terms of single-electron spin-orbitals.
\cite{SLAT60} 
\begin{eqnarray}
\sum_i \langle \Phi_{\mu} |f_i| \Phi_{\nu} \rangle &=& \frac{1}{i} \sum_i 
\int d^3 r_1 \cdots d^3 r_N \sum_{\sigma_1, \cdots, \sigma_N}
\psi_{1 \sigma_1}^{\ast}(\vec{r}_1) \chi_{\sigma_1}^{\ast}
\psi_{2 \sigma_2}^{\ast}(\vec{r}_2) \chi_{\sigma_2}^{\ast} \cdots
\psi_{N \sigma_N}^{\ast}(\vec{r}_N) \chi_{\sigma_N}^{\ast} \nonumber \\
& & \cdot
\sum_{\nu} \frac{Z_{\nu}}{2c^2} \vec{S}_i \cdot 
\left( \vec{\nabla}_i \times  \frac{(\vec{r}_i - \vec{R}_{\nu})}
{|\vec{r}_i - \vec{R}_{\nu}|^3} \right) 
\left| \begin{array}{ccc}
       \psi^{\prime}_{1 \sigma_1}(\vec{r}_1) \chi^{\prime}_{\sigma_1} & \cdots & 
       \psi^{\prime}_{1 \sigma_1}(\vec{r}_N) \chi^{\prime}_{\sigma_1} \\
       \vdots & \ddots & \vdots \\
       \psi^{\prime}_{N \sigma_N}(\vec{r}_1) \chi^{\prime}_{\sigma_N} & \cdots &
       \psi^{\prime}_{N \sigma_N}(\vec{r}_N) \chi^{\prime}_{\sigma_N} 
       \end{array}
\right| 
\label{eq:vls-7}
\end{eqnarray}
where $\vec{\nabla}_i$ is also applied to the single-electron 
spin-orbitals and the normalization factor $1/N!$ cancels out
$N!$ permutations of the $N$ single-electron spin-orbitals
in $\langle \Phi_{\mu} |$. 
As can be seen from Eq.~(\ref{eq:vls-7}), because all spin-orbitals
$\psi_{1 \sigma_1}\chi_{\sigma_1}$,$\cdots$
$\psi_{N \sigma_N}\chi_{\sigma_N}$,
$\psi^{\prime}_{1 \sigma_1}\chi^{\prime}_{\sigma_1}$,$\cdots$,
$\psi^{\prime}_{N \sigma_N}\chi^{\prime}_{\sigma_N}$,
are orthonormal to each other,
the off-diagonal elements of the one-electron operator $f_i$ 
are nonvanishing when only one single-electron
spin-orbital is different between $|\Phi_{\mu} \rangle$ 
and $|\Phi_{\nu} \rangle$. The off-diagonal elements of the 
two-electron operator $g_{ij}$ are nonvanishing when no more than 
two single-electron spin-orbitals are different between 
$|\Phi_{\mu} \rangle$ and $|\Phi_{\nu} \rangle$. Different 
single Slater determinants (or different low-energy spin excitations) 
are not connected by the one- or two-electron operators, so 
$\langle \Phi_{\mu} |V_{LS}| \Phi_{\nu} \rangle = 0$ for $\mu \neq \nu$.
Hence, the off-diagonal contributions to the energy shift
$\Delta_2$ is zero. As shown in Table~\ref{table:1}, the
calculated magnetic anisotropy barriers for the low-energy spin
excitations are almost the same as that for the ground state
$S=10$ manifold. This indicates that the diagonal elements of the
spin-orbit interaction [Eq.~(\ref{eq:vls-6})] do not vary much
with different spin excitations: $\langle \Phi_{\mu} |V_{LS}|
\Phi_{\mu}  \rangle \sim \langle \Phi_{\nu} |V_{LS}| \Phi_{\nu}
\rangle$ for $\mu\neq \nu$. Then $\langle \Phi | V_{LS} | \Phi
\rangle \sim \langle \Phi_{\mu} |V_{LS}| \Phi_{\mu}  \rangle (
\sum_{\mu} |C_{\mu}|^2 )$ where $\sum_{\mu} |C_{\mu}|^2=1$ by
definition. So the magnetic anisotropy barriers for low-lying 
excited-state manifolds must be approximately the same as that for 
the ground-state manifold. Note that we have calculated second-order
anisotropy barriers only so that the barriers shown in Table~\ref{table:1}
are all between 54~K and 55~K.
The results of this discussion suggest that the DFT-based determination
of MAE is valid under the following conditions: (i) single-electron and two-electron 
excitations cost an order of eV, (ii) considered collinear spin excitations
cost an order of magnitude less energy than the single-electron and two-electron 
excitations.

\section{Conclusion}

We have calculated the intramolecular exchange interactions 
among Mn spins in the single-molecule magnet Mn$_{12}$ 
([Mn$_{12}$O$_{12}$(HCOO)$_{16}$(H$_2$O)$_4$]) 
considering localized low-energy spin excitations using density-functional 
theory. Our calculated values of the exchange constants agree with the results from
Refs.~\onlinecite{REGN02,BOUK02}. With our calculated exchange constants, 
we have diagonalized the isotropic Heisenberg Hamiltonian using the Lanczos method. 
We have confirmed that the ground state is $S=10$ and that there 
is a $S=9$ first-excited manifold located at 40.5~K above
the $S=10$ ground-state manifold. This energy gap (40.5~K) is an upper limit
because it is likely reduced by including more exact treatments of the Coulomb 
potential or electron correlations. This indicates that the first excited
manifold may be situated much lower than the high energy levels within the
$S=10$ ground-state manifold. We have also calculated the magnetic anisotropy
barriers for the low-energy spin excitations using DFT and the
second-order perturbation theory with the assumption that the spin-orbit
coupling is the most important interaction. Our DFT calculations have showed that
the anisotropy barriers for the low-energy spin excitations are 
approximately the same as that for the $S=10$ ground state. 
From this result and our understanding of the spin-orbit interaction,
we conclude that the anisotropy barriers for the low-lying excited
manifolds are approximately same as that for the $S=10$ ground state.

\section*{Acknowledgments}
KP was funded by W. M. Keck Foundation, MRP and CSH were supported
in part by ONR and the DoD HPC CHSSI program.

\clearpage

\begin{table}
\begin{center}
\caption{Here $M_s=6-c$ denotes $M_s=6$,
where $M_s$ is an eigenvalue of the total spin operator projected 
along the easy axis and ``c'' is attached to distinguish between
different types of $M_s=6$ states. From the second column, shown are
flipped spins to create low-energy spin excitations
relative to the $S=10$ ground-state
labeled in Fig.\protect{\ref{fig:mn12sch}}, Ising energy expressions,
DFT-calculated energies relative to the ground state, $\Delta E$
(in units of eV),  the energy differences between DFT results
and least-square-fit (LSF), E$^{\mathrm DFT}$$-$E$^{\mathrm LSF}$, 
and the magnetic anisotropy barriers (MAE) (in units of K).  }
\label{table:1}
\begin{ruledtabular}
\begin{tabular}{cccccc}
$M_s$ & flipped spins & Ising Energy & $\Delta E$ (eV) 
& E$^{\mathrm DFT}$$-$E$^{\mathrm LSF}$ (eV) & MAE (K) \\ \hline
10  &  & $E_0-24 J_1 -48 J_2 + 27 J_3 + 64 J_4$ & 0 & -0.0049 & 54.1  \\ \hline
6-c & 9 & $E_0-12 J_1 -48 J_2 + 27 J_3 + 32 J_4$ & 0.0352 & -0.0064 & 54.8 \\  \hline
9-b & 1, 9 & $E_0-24 J_1 -24 J_2 + 36 J_4$ & 0.0598 & -0.0134 & 54.5 \\ \hline
6-b & 5 & $E_0-24 J_1 -24 J_2 + 27 J_3 + 32 J_4$ & 0.0780 & 0.0095 & 55.2 \\ \hline
5-b & 1, 5, 9 & $E_0-24 J_1 -24 J_2 + 32 J_4$ & 0.0902 & 0.0171 & 54.8 \\ \hline
5-a & 1, 7, 9 & $E_0-24 J_1$ & 0.1328 & -0.0045 & 55.4 \\ \hline
8-b & 1, 4, 5, 9 & $E_0-12 J_1 -24 J_2 -9 J_3 + 36 J_4$ & 0.1361 & 0.0014 & 54.1  \\ \hline
8 & 1, 4, 9, 12 & $E_0-24 J_1 -9 J_3$ & 0.1377 & -0.0012 & 54.9 \\ \hline
9-c & 4, 5 & $E_0-12 J_1 -24 J_2 +36 J_4$ & 0.1445 & 0.0114 & 54.9  \\  \hline
13 & 1 & $E_0-12 J_1 -24 J_2 +64 J_4$ & 0.1496 & -0.0064 & 53.6 \\ \hline
8-c & 1, 8, 9 & $E_0-12 J_1 -9 J_3$ & 0.1929 & -0.0057 & 54.8 \\ \hline
9-d & 4, 9 & $E_0-24 J_2 +32 J_4$ & 0.1955 & 0.0027 & 55.0 \\
\end{tabular}
\end{ruledtabular}
\end{center}
\end{table}

\begin{table}
\begin{center}
\caption{Comparison of our DFT-calculated intramolecular
exchange couplings with those from LDA+U calculations\cite{BOUK02}, 
diagonalization of the Heisenberg model\cite{RAGH01,REGN02},
and experimental results\cite{SESS93} in units of K.
A positive sign denotes an antiferromagnetic coupling. }
\label{table:2}
\begin{ruledtabular}
\begin{tabular}{cccccc}
  &   DFT  & LDA+U[Ref.\onlinecite{BOUK02}] & Ref.\onlinecite{REGN02}
& Ref.\onlinecite{RAGH01} & Exp.[Ref.\onlinecite{SESS93}]  \\ \hline
$J_1$ & 115  & 94 & 119  &  430   &  432 \\ \hline
$J_2$ &  84  & 52 & 118  &  170   &  173 \\ \hline
$J_3$ & $-4$ & 60 & $-8$ &  170   &  166 \\ \hline
$J_4$ &  17  & 14 &  23  & $-129$ &    0 \\
\end{tabular}
\end{ruledtabular}
\end{center}
\end{table}

\begin{table}
\begin{center}
\caption{Several lowest-energy eigenvalues of the Hamiltonians for
a total number of $N$ $M_s=-12$, $M_s=-11$, $M_s=-10$, $M_s=-9$, and $M_s=-8$ states.} 
\label{table:3}
\begin{ruledtabular}
\begin{tabular}{crccccc}
$M_s$ & $N$  & Energy (K)\\ \hline
$-12$ & 269148 & -2574.2988 \\ \hline
$-11$ & 484144 & -3034.6061 \\ 
      &        & -2926.5083 \\ \hline
$-10$ & 817176 & -3464.9569 \\ 
      &        & -3034.6061 \\ \hline
$-9$ & 1299632 & -3464.9569 \\ 
     &         & -3424.4283  \\
     &         & -3417.4348 \\ 
     &         & -3384.2171 \\ \hline
$-8$ & 1954108 & -3464.9569 \\
     &         & -3424.4283 \\
     &         & -3417.4348 \\ 
     &         & -3384.2171 \\  
\end{tabular}
\end{ruledtabular}
\end{center}
\end{table}

\clearpage

\end{document}